\documentclass[twocolumn, longbibliography, prr, amsmath, amssymb,superscriptaddress]{revtex4-2}
\usepackage{amsfonts}
\usepackage{graphicx}
\usepackage[colorlinks,linkcolor=blue,citecolor=blue,urlcolor=blue]{hyperref}
\usepackage{float}
\usepackage[normalem]{ulem}

\begin{document}
\title{Stochastic light in a cavity: A Brownian particle in a scalar potential?}

\author{J. Busink}
\affiliation {Center for Nanophotonics, AMOLF, Science Park 104, 1098 XG Amsterdam, The Netherlands}

\author{P. Ackermans}
\affiliation {Center for Nanophotonics, AMOLF, Science Park 104, 1098 XG Amsterdam, The Netherlands}

\author{K. G. Cogn\'{e}e}
\affiliation {CDI, City College of New York, New York, NY 10031, USA}

\author{S. R. K. Rodriguez} \email{s.rodriguez@amolf.nl}
\affiliation {Center for Nanophotonics, AMOLF, Science Park 104, 1098 XG Amsterdam, The Netherlands}

\begin{abstract}
{The non-equilibrium dynamics of stochastic light in a coherently-driven nonlinear cavity resembles the equilibrium dynamics of a Brownian particle in a scalar potential. This resemblance has been known for decades, but the correspondence between the two systems has never been properly assessed. Here we demonstrate that this correspondence can be exact, approximate, or break down, depending on the cavity nonlinear response and driving frequency. For weak on-resonance driving, the nonlinearity vanishes and the correspondence is exact: The cavity dissipation and driving amplitude define a scalar potential, the noise variance defines an effective temperature, and the intra-cavity field satisfies Boltzmann statistics. For moderately strong non-resonant driving, the correspondence is approximate:  We introduce a potential that approximately captures the nonlinear dynamics of the intra-cavity field, and we quantify the accuracy of this approximation via deviations from Boltzmann statistics.  For very strong non-resonant driving, the correspondence breaks down:  The intra-cavity field dynamics is governed by non-conservative forces which preclude a description based on a scalar potential only.  We furthermore show that this breakdown is accompanied by a phase transition for the intra-cavity field fluctuations, reminiscent of a non-Hermitian phase transition. Our work establishes clear connections between optical and stochastic thermodynamic systems, and suggests that many fundamental results for overdamped Langevin oscillators may be used to understand and improve resonant optical technologies. }
\end{abstract}
\date{\today}
\maketitle

\section{Introduction}
Many advances in optical physics have resulted from identifying a correspondence between non-equilibrium behavior of light and equilibrium behavior of matter. For instance, Haken realized that the onset of lasing corresponds to a second order phase transition in equilibrium~\cite{Haken70}. He furthermore identified deep connections between optics and Ginzburg-Landau theory~\cite{Haken70, Haken75}, and thereby pioneered research on phase transitions of photons. This research has flourished recently, resulting for example in the discovery of novel dissipative phase transitions~\cite{Fitzpatrick17, Fink17, Rodriguez17, Casteels17, Biondi17, Fink18, Young20} and applications to quantum technologies~\cite{Hartmann16, Mendoza16, Noh16, Heugel19}. A more recent example is due to Foss-Feig and co-workers, who realized that an array of bistable optical resonators admits an effective equilibrium description in terms of a classical Ising model~\cite{Foss17}. This correspondence is promising for solving non-deterministic polynomial time (NP)-hard problems~\cite{Lucas14, Liew19}, for which no efficient algorithm exists~\cite{Barahona82}.

\begin{figure}[b]
	\includegraphics[width=\columnwidth]{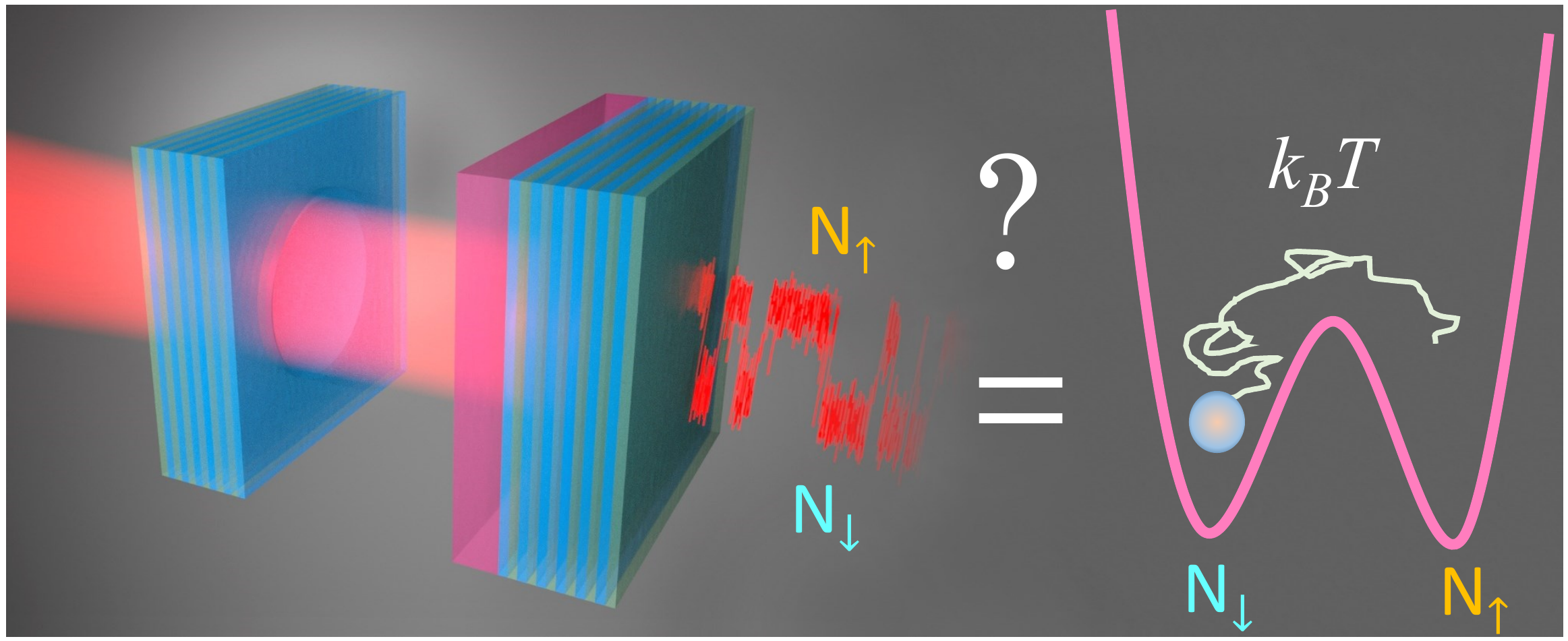}
	\caption{\label{fig:1} Left: The transmission of a coherently-driven nonlinear optical cavity switches between two states due to the influence of noise. Right: A Brownian particle in a scalar double well potential. The figure, overall, illustrates the main question motivating this manuscript: Is stochastic light in a coherently-driven cavity a Brownian particle in a scalar potential?}
\end{figure}

In the 1980's, Risken and co-workers made an interesting analogy between a bistable optical cavity and a Brownian particle in a double well potential~\cite{Risken87, Vogel89, Risken90}. They associated bistable optical states with the minima of the potential, and fluctuations of the intra-cavity light field with the thermal motion of the particle. Despite the long history of this analogy, its exact or approximate validity has never been properly assessed. Recently, Andersen and co-workers defined a metapotential for a bistable resonator~\cite{Andersen20}. Using this metapotential and the equilibrium theory of Kramers~\cite{Kramers40}, they approximately reproduced the system's dynamics in certain parameter regimes. In other regimes, inconsistencies with quantum theory were attributed to quantum effects rather than to the questionable validity of the metapotential. This prompts the questions: Can a scalar potential be defined for an optical resonator? And how far does the correspondence to equilibrium physics go? Figure~\ref{fig:1} illustrates the essence of these questions, which are the motivation for this work.

Here we demonstrate that the correspondence between stochastic light in a coherently-driven cavity and a Brownian particle in a scalar potential can be  exact, approximate, or break down, depending on the parameters of the optical system. In Section \ref{section:model} we introduce the model for an optical cavity, and show that the complex intra-cavity light field is mathematically equivalent to two overdamped Langevin oscillators. In Section \ref{section:linear} we show that if (and only if) the cavity is linear and driven on resonance, the two Langevin oscillators decouple and a scalar potential fully captures the light field dynamics. For arbitrarily strong and non-resonant driving, the oscillators exert non-conservative forces upon each other. This precludes describing the optical cavity in terms of a scalar potential only. Nonetheless, in Section \ref{section:NL} we show that an approximate potential can still be defined in certain parameter regimes. We then quantify the validity of this potential as a function of the driving conditions in Section \ref{section:deviation}, and reveal how the stability of the fixed points determines the validity of our potential in Section \ref{section:fluctuations}. The two models we relate --- the overdamped Langevin oscillator and a single-mode coherently-driven cavity --- are cornerstones of stochastic thermodynamics on one hand and resonant optics on the other hand. While connections between classical (deterministic) thermodynamics and optics have been known for decades,  our work points to a new frontier of physics at the intersection of \textit{stochastic} thermodynamics~\cite{Sekimoto98, Jarzynski11, Seifert12, Ciliberto17} and resonant optics. Section \ref{section:conclusion} presents our perspective towards that frontier and a summary of our results.

\section{The model} \label{section:model}

Consider a coherently-driven single-mode cavity with Kerr nonlinearity. Within the truncated Wigner approximation and in a frame rotating at the driving frequency $\omega$, the intra-cavity light field $\alpha$ satisfies the following equation of motion~\cite{CarusottoRMP_2013}:

    \begin{equation}
        i\Dot{\alpha}=\left(-\Delta-i\frac{\Gamma}{2}+U|\alpha|^2\right)\alpha+i\sqrt{\kappa_L}A+D\zeta(t).
        \label{eq: 1}
    \end{equation}

\noindent $\Delta=\omega-\omega_0$ is the detuning of the resonance frequency $\omega_0$ from $\omega$. $\Gamma = \gamma_a +\kappa_L+\kappa_R$ is the total loss rate, with $\gamma_a$ the absorption rate and $\kappa_{L,R}$ the input-output rate through the left or right mirror. $U$ is the Kerr nonlinearity strength. $A$ is the amplitude of the coherent driving field. $\zeta(t)=\zeta_R(t) + i\zeta_I(t)$ provides Gaussian white noise in the real and imaginary parts of the light field. $\zeta_{R,I}$ each have zero mean ($\langle\zeta_R(t)\rangle =\langle\zeta_I(t)\rangle=0$), and are delta-correlated ($\langle\zeta_R(t')\zeta_R(t)\rangle= \langle\zeta_I(t')\zeta_I(t)\rangle= \delta(t'-t)$). Moreover, $\zeta_{R}$ and $\zeta_{I}$ are mutually uncorrelated. The standard deviation of each noise field is $D$.

We begin our analysis by decomposing Equation~\ref{eq: 1} into real and imaginary parts. Defining $\alpha=\alpha_R+i\alpha_I$ and $\Omega = UN -\Delta$ with $N=|\alpha|^2$ the photon number, we get
    \begin{align}
  \begin{pmatrix}\Dot{\alpha_R} \\ \Dot{\alpha_I}\end{pmatrix} =
       \underbrace{   \begin{pmatrix}
   -\frac{\Gamma}{2} & \Omega \\
     -\Omega &-\frac{\Gamma}{2}
   \end{pmatrix}
   \begin{pmatrix}
    \alpha_R\\ \alpha_I \end{pmatrix}  + \begin{pmatrix}
    \sqrt{\kappa_L}A\\ 0 \end{pmatrix}}_{F}+ D\begin{pmatrix}
    \zeta_R (t)\\ \zeta_I (t) \end{pmatrix}.
\label{eq: 2}
\end{align}

\noindent The decomposition reveals that a single stochastic Kerr-nonlinear cavity is mathematically equivalent to two coupled overdamped Langevin oscillators. The real and imaginary parts of the light field, $\alpha_{R,I}$, represent the displacement from equilibrium of the oscillators. The oscillators evolve under the influence of a deterministic force $F$ and a stochastic force $D\zeta$. The oscillators are coupled by the off-diagonal elements of the first matrix in the right hand side of Equation~\ref{eq: 2}.

To determine if a scalar potential $V = -\int\vec{F}d\vec{\alpha}$ can be defined for our cavity, recall that $F$ must be conservative and irrotational for $V$ to exist. The magnitude of the curl of $F$ for our cavity is
\begin{align}
    \left| \vec{\nabla}\times\vec{F}  \right| = \left| \bigg{(}\frac{\partial \Dot{\alpha}_I}{\partial \alpha_R}-\frac{\partial \Dot{\alpha}_R}{\partial \alpha_I}\bigg{)} \right| =2\Omega.
\end{align}
\noindent Hence, if an only if $\Omega = 0$, $\vec{F}$ is irrotational and $V$ exists.

Our analysis can be generalized to systems of coupled oscillators (cavities) using a classic result of graph theory: For $V$ to exist,  the adjacency matrix $\mathcal{A}$ must be symmetric~\cite{Manoel15, Aguiar19}. For our single-mode cavity, the adjacency matrix is

 \begin{align}
  \mathcal{A} =
 \begin{pmatrix}
   0 &\Omega \\
   -\Omega &0
   \end{pmatrix} ,
\end{align}

\noindent which is anti-symmetric whenever $\Omega \neq 0$. Therefore, we can only expect to fully capture the dynamics of light in an optical cavity with a scalar potential $V$ when the response is linear ($U=0$) and the driving is on resonance ($\Delta=0$), such that $\Omega = 0$.

\section{Exact Potential for a Linear Cavity Driven on Resonance} \label{section:linear}

\begin{figure}
	\includegraphics[width=\columnwidth]{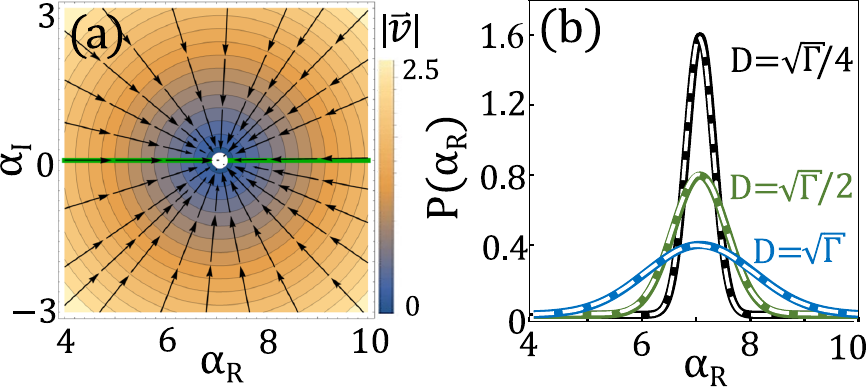}
	\caption{\label{fig:2}(a) Phase portrait of a linear cavity driven on resonance. $\alpha_R$ and $\alpha_I$ are the real and imaginary parts of the intra-cavity field $\alpha$, respectively. $\vec{v}= (\Dot{\alpha}_R,\Dot{\alpha}_I)$ is the force acting on the intra-cavity field. Arrows and color represent the direction and magnitude of the force, respectively. The white dot is the sole stable fixed point. (b) PDF of $\alpha_R$ for three values of the standard deviation of the noise $D$ relative to the dissipation $\Gamma$. Solid curves are results from numerical simulations using Equation~\ref{eq: 1}. Dashed white curves are Boltzmann distributions. The potential $V_R$ for the Boltzmann distributions is defined along the green line in (a), where $\Dot{\alpha}_I=0$. The effective temperature is given by $T=D^2/2k_B$, with $k_B$ the Boltzmann constant.}
\end{figure}

Let us assume that the optical cavity is strictly linear ($U=0$) and driven on resonance ($\Delta=0$). In this case, $\Omega =0$, the two oscillators decouple, and the non-conservative force vanishes. Neglecting noise ($D=0$), Fig.~\ref{fig:2}(a) shows the phase portrait of the system. We plot the local time-evolution of the field $\vec{v}= (\Dot{\alpha}_R,\Dot{\alpha}_I)$ as black arrows, and its magnitude $|\vec{v}|$ in color. In general, $\vec{v}$ represents the total deterministic force locally experienced by light in the $\alpha_R \alpha_I$-plane. Notice in Fig.~\ref{fig:2}(a) that the vectors $\vec{v}$ are perpendicular to contours of constant force $|\vec{v}|$ and directed towards the minimum in $|\vec{v}|$. This is the typical behavior of a gradient flow system. Indeed, for $\Omega =0$, Equation~\ref{eq: 2} reduces to a set of two decoupled overdamped oscillators each subject to a gradient (conservative) force. If we now allow $D \neq 0$, the equations of motion for these decoupled oscillators are:

\begin{align}
    \Dot{\alpha}_{R,I} = -\frac{\partial V_{R,I}}{\partial \alpha_{R,I}}+D\zeta_{R,I}.
    \label{eq: 3}
\end{align}
\noindent The potential functions in Eq.~\ref{eq: 3} are

\begin{subequations} \label{eq: 4}
\begin{align}
V_R &= \frac{\Gamma}{4}\alpha_R^2-\sqrt{\kappa_L}A\alpha_R, \label{eq: 4a}    \\
V_I &= \frac{\Gamma}{4}\alpha_I^2 .
\end{align}
\end{subequations}

\noindent Equations~\ref{eq: 3} and~\ref{eq: 4} show that the potential for both overdamped oscillators is harmonic. The only difference between the oscillators is that the equilibrium position for $\alpha_R$ is displaced from zero
in proportion to $\sqrt{\kappa_L}A$, which is the laser amplitude entering the cavity.

Next we illustrate the statistical properties of stochastic light in the cavity. For this purpose, we performed stochastic calculations of Equation~\ref{eq: 1} using the xSPDE Matlab toolbox~\cite{xSPDE}. Figure~\ref{fig:2}(b) shows the probability density function for $\alpha_{R}$, $P(\alpha_{R})$, for three values of $D/\sqrt{\Gamma}$ as curves of different color. Figure~\ref{fig:2}(b) also shows, as white dashed curves, the equilibrium Boltzmann distribution obtained for a Brownian particle in the scalar potential $V_R$. Concretely, we plot

\begin{align}
    P(\alpha_{R}) = \mathcal{N}e^{-V_{R}/k_bT},
    \label{eq: 5}
\end{align}

\noindent with $k_b$ the Boltzmann constant and $\mathcal{N}$ a normalization constant. The temperature $T$ is fixed by the noise variance $D^2$ according to $T=D^2/2k_B$. Notice in Fig.~\ref{fig:2}(b) how $P(\alpha_{R})$ spreads as $D/\sqrt{\Gamma}$ increases, in perfect agreement with the Boltzmann distribution for a gas with increasing temperature. We stress that the Boltzmann distributions $P(\alpha_{R})$ in Fig.~\ref{fig:2}(b) are first-principles calculations and not fits to the numerical data. We inserted the potential $V_{R}$ from Eq.~\ref{eq: 4a} into Eq.~\ref{eq: 5} to calculate $P(\alpha_{R})$. $V_{I}$ can be neglected because $\alpha_{I}$ is decoupled from $\alpha_{R}$, and the driving field acts on $\alpha_{R}$ only.

The preceding analysis demonstrates that a linear optical cavity driven on resonance is mathematically equivalent to an overdamped Langevin oscillator in equilibrium. Through this powerful correspondence, we can use the theoretical framework of statistical physics for understanding and optimizing resonant optical systems. A critical question remains: What is the meaning of the `temperature' of the light field in the cavity? Clearly, that effective temperature is unrelated to the temperature of the medium inside the cavity. In fact, the effective temperature of the light field can be  externally controlled by imprinting noise on the driving laser using modulators~\cite{Abbaspour14, Peters21} and without changing the cavity dissipation. We therefore propose that the effective temperature of the light field should be understood from the perspective of the kinetic theory of gases. From that perspective, the temperature of an ideal gas is related to the average kinetic energy of the particles. Simply put, temperature is motion. Indeed, a higher temperature increases the probability of finding a particle away from its equilibrium position at zero temperature. This is exactly what noise in the laser amplitude and phase does to the intracavity light field: it increases the probability of finding field amplitudes and phases away from the equilibrium value at zero noise. In the remainder of this manuscript, we avoid further discussions about the `meaning' of the exact mathematical correspondence discussed above. Instead, we introduce and assess an approximate potential for a nonlinear cavity driven out of resonance.

\begin{figure}[t]
	\includegraphics[width=\columnwidth]{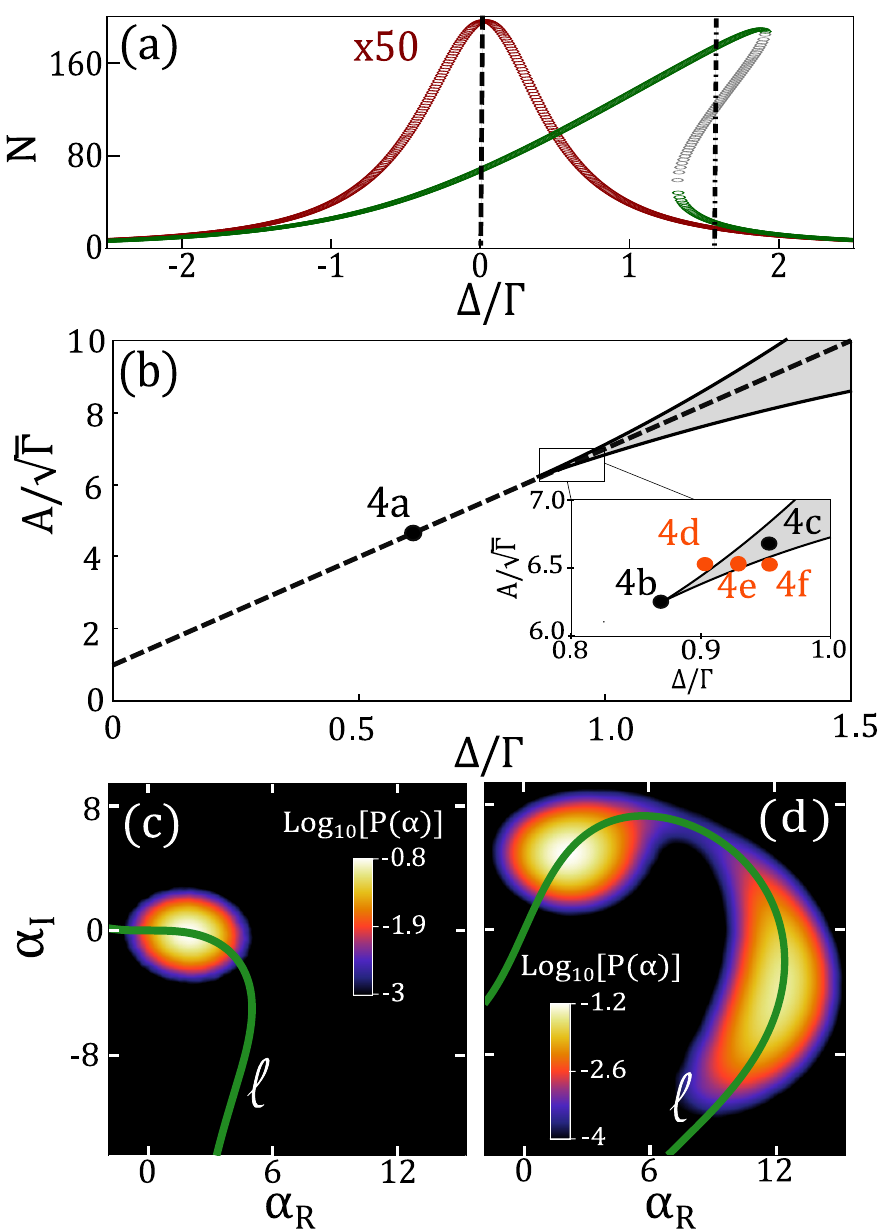}
 	\caption{\label{fig:3}(a) Number of photons $N$ as a function of the detuning $\Delta$ referenced to the dissipation $\Gamma$, for two different driving amplitudes $A$. $A/\sqrt{\Gamma}=1.4$ for the red curve, where all states are stable. $A/\sqrt{\Gamma}=9.71$ for the green and gray curves, which correspond to stable and unstable states, respectively. Dashed and dashed-dotted lines indicate the detunings considered in (c) and (d), respectively. (b) Phase diagram of a Kerr-nonlinear cavity, indicating the number of stable steady states versus driving parameters. White and gray areas correspond to one and two stable states, respectively. Entering the bistability along the dashed line corresponds to a supercritical pitchfork bifurcation. The inset zooms into the vicinity of the critical point, and indicates the driving conditions considered in Fig.~\ref{fig:4}. (c,d) PDF for the intra-cavity field obtained by numerically solving Equation ~\ref{eq: 1}. The green curves indicate the path $\ell$ where $\Dot{\alpha}_I=0$ and the potential $V_{\mathrm{app}}$ is defined.}
\end{figure}

\section{Approximate Potential for a Nonlinear Cavity} \label{section:NL}

We have previously shown that the distribution $P(\alpha_{R})$ perfectly agrees with the Boltzmann distribution when the cavity response is strictly linear (i.e., $U=0$) and the driving is on resonance (i.e., $\Delta=0$). A similar agreement is expected for $U \neq 0$ when the driving amplitude $A$ is arbitrarily small. Such a weak driving amplitude ensures that $UN \ll \Gamma$ and the oscillators $\alpha_R$ and $\alpha_I$ effectively decouple, provided that $\Delta=0$. In this and the following sections, we pursue an understanding of the physics when $U$ and $A$ are sufficiently large for $UN$ to be commensurate with $\Gamma$. We set $U/\Gamma=0.01$ for all calculations in the remaining of the manuscript, and vary  $A$ and $\Delta$.

In Fig.~\ref{fig:3}(a) we illustrate how a finite nonlinearity affects the spectrum of a deterministic ($D=0$) cavity as the driving amplitude $A$ increases. For weak driving, the red curve shows an approximately Lorentzian resonance lineshape. For strong driving, the green and gray curves correspond to stable and unstable states. The resonance lineshape bends towards positive $\Delta$ because $U>0$, and a region of bistability emerges.

Figure~\ref{fig:3}(b) depicts as white and gray areas the parameter range for which one can observe respectively one or two stable steady states.  Figures~\ref{fig:3}(c,d) show probability density functions (PDFs) for the complex intra-cavity light field $\alpha$ at two distinct driving conditions. These PDFs were calculated based on stochastic trajectories of $\alpha(t)$. We calculated 8 trajectories with different realization of the noise $\zeta(t)$, all with a large duration $\Gamma t=10^6$ and a standard deviation of the noise $D/\sqrt{\Gamma}=1$. Figure~\ref{fig:3}(c) was obtained for $A/\sqrt{\Gamma}=1.4$ and $\Delta/\Gamma = 0$, which probes the state at the intersection of the red curve and the dashed line in Fig.~\ref{fig:3}(a). The slightly larger uncertainty of the state along $\alpha_R$ than along $\alpha_I$ is a mild squeezing effect due to the nonlinearity. Figure~\ref{fig:3}(d) shows the PDF for $A/\sqrt{\Gamma}=9.71$ and $\Delta/\Gamma = 1.5$, which probes the states at the intersections of the green curves and the dash-dotted line in Fig.~\ref{fig:3}(a). The observed bimodal distribution indicates bistability.

Strictly speaking, the dynamics of a nonlinear cavity cannot be fully described by a scalar potential. This is due to the fact that $\Omega\neq0$, which means that the two oscillators $\alpha_{R,I}$ are coupled and exert a non-conservative force upon each other.  However, for sufficiently weak coupling, i.e.,  $\Omega \ll \Gamma$, the dynamics of the undriven oscillator may be disregarded. In that case, an approximate potential $V_{\mathrm{app}}$ may capture the essential dynamics of the full system. To test this idea, we plot the values of ($\alpha_{R}$, $\alpha_{I}$) for which $\Dot{\alpha}_I = 0$ as green curves in Figs.~\ref{fig:3}(c,d). Along this one-dimensional path $\ell$, the deterministic force on the undriven oscillator is zero. Notice how the path $\ell$ passes through the main features of the PDF even in the nonlinear regime. Remarkably, $\ell$ closely follows the most probable path between the two bistable states in Fig.~\ref{fig:3}(d), as evidenced by the region of maximum probability connecting the two peaks in the PDF. Based on this observation, we propose defining $V_{\mathrm{app}}$ along $\ell$.

\begin{figure}[t]
	\includegraphics[width=\columnwidth]{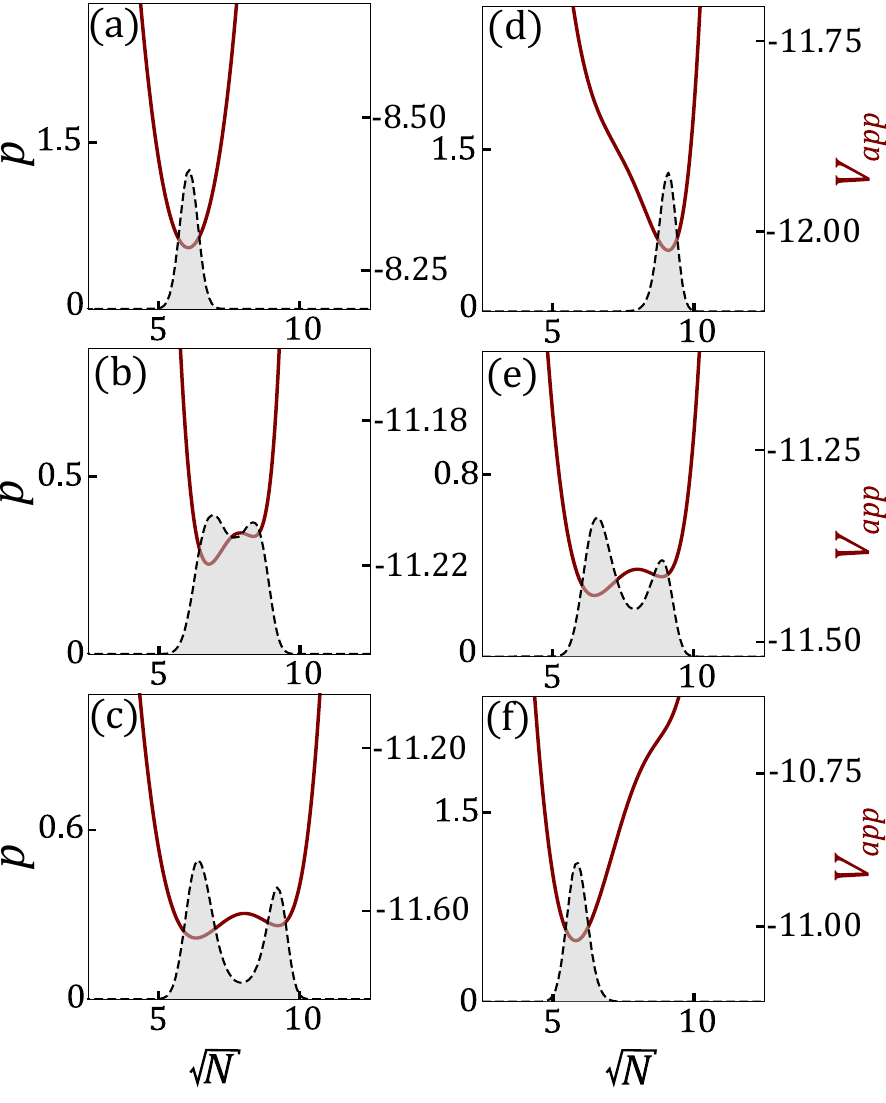}
	\caption{\label{fig:4} (a-f) Red curves are the potential $V_{\mathrm{app}}$. Shaded areas are PDFs for the field amplitude $\sqrt{N}$. (a-c): Driving conditions are along the dashed line in Fig.~\ref{fig:3}(b). The PDF and $V_{\mathrm{app}}$ are reshaped as $A/\sqrt{\Gamma}$ and $\Delta/\Gamma$ increase and the system undergoes a supercritical pitchfork bifurcation. (d-f): Driving conditions are indicated by the orange dots in Fig.~\ref{fig:3}(b). The PDF and $V_{\mathrm{app}}$ tilt due to a change in $\Delta/\Gamma$. }
\end{figure}

Along the path $\ell$, the time evolution of the driven oscillator acts as a local force approximately capturing the full system's dynamics:

\begin{align}
    F_{\mathrm{app}} = \Dot{\alpha}_R\bigg{\rvert}_{\Dot{\alpha}_I=0}
    = \sqrt{\kappa_L}A -\frac{\Gamma}{2\alpha_R}(\alpha_I^2+\alpha_R^2).
    \label{eq: 6}
\end{align}
The approximate potential $V_{\mathrm{app}}$ on this path is obtained by integrating $F_{\mathrm{app}}$ along $\ell$:
\begin{align}
  V_{\mathrm{app}}(\alpha_R,\alpha_I) = -\int_{\ell}  F_{\mathrm{app}}d\sqrt{N},
  \label{eq: 7}
\end{align}
\noindent with $\sqrt{N}=\sqrt{(\alpha_I^2+\alpha_R^2)}$ the variable of integration. We take $\sqrt{N}$ (instead of $N$) as the variable of integration for two reasons. First, this choice ensures $V_{\mathrm{app}}$ has units of energy, since $F_{\mathrm{app}}d\sqrt{N}$ can be understood as a force times displacement, which has units of energy. Second, our choice ensures $V_{\mathrm{app}}$ has the same units as $V_{R}$, since $V_{\mathrm{app}} \rightarrow V_{R}$ for $\Omega \rightarrow 0$.

Figure~\ref{fig:4} shows $V_{\mathrm{app}}(\sqrt{N})$ as red curves for different driving conditions. Figures~\ref{fig:4}(a,b,c) are evaluated along the dashed line in Fig.~\ref{fig:3}(b), crossing the critical point $\{\Delta_c, A_c\} = \{\Gamma\sqrt{3}/2,\Gamma^{3/2}3^{-3/4}/\sqrt{\kappa_L U}\}$~\cite{Foss17}. The observed transformation of $V_{\mathrm{app}}$ from single well to double well corresponds to a system undergoing a supercritical pitchfork bifurcation. To demonstrate how $V_{\mathrm{app}}$ captures the full system dynamics, the shaded areas in Fig.~\ref{fig:4} show PDFs obtained from stochastic simulations of Equation~\ref{eq: 1}. Notice the good agreement between the peaks in the PDF and and the dips in $V_{\mathrm{app}}$ for the various parameter values. A more quantitative comparison is reserved for the next section.

Figures~\ref{fig:4}(d,e,f) show that $V_{\mathrm{app}}$ approximately captures the distribution of light in the cavity also when $\Delta/\Gamma$ is varied while $A/\sqrt{\Gamma}$ is constant. In particular, we plot $V_{\mathrm{app}}$ at the driving conditions indicated by the orange dots in Fig.~\ref{fig:3}(b). Figures~\ref{fig:4}(d,e,f) show how $V_{\mathrm{app}}$ tilts from one side to another as $\Delta/\Gamma$ is varied. Correspondingly, the numerically calculated PDFs for the full system (Equation~\ref{eq: 1}) show the same tilting behavior. This evidences that $V_{\mathrm{app}}$ successfully captures the essential physics, at least qualitatively.

\section{Deviations from Boltzmann statistics} \label{section:deviation}

\begin{figure}[t]
	\includegraphics[width=\columnwidth]{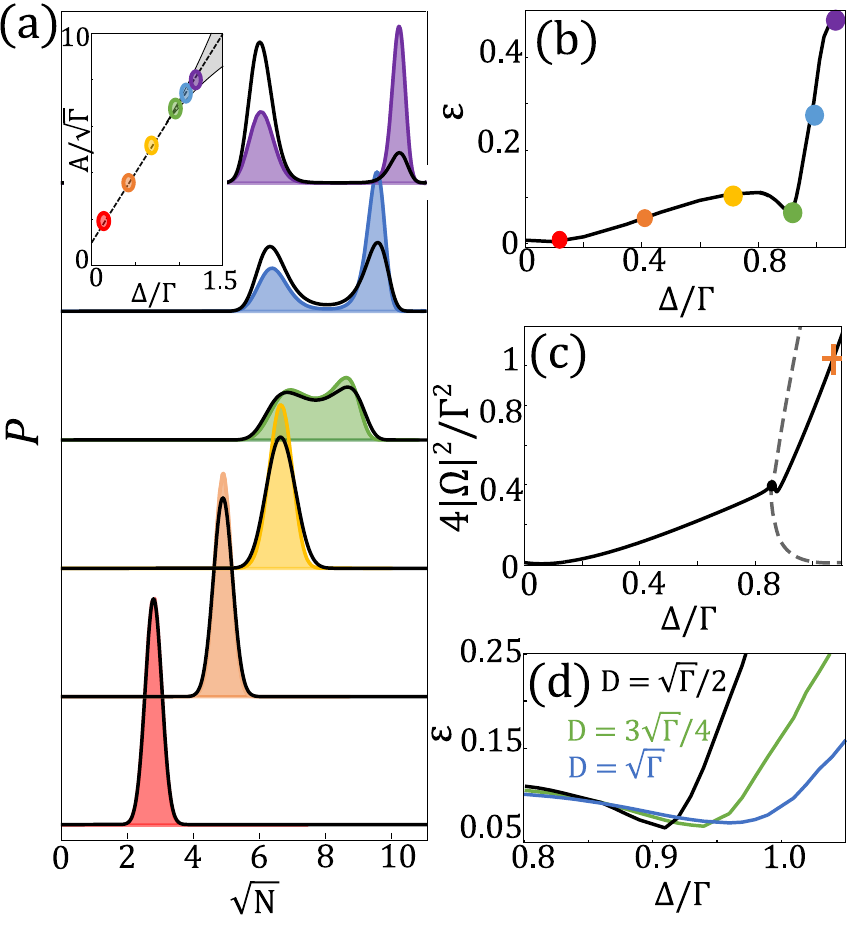}
	\caption{\label{fig:5} (a) PDF of the field amplitude $\sqrt{N}$ for 6 values of the driving amplitude $A/\sqrt{\Gamma}$ and the detuning $\Delta/\Gamma$, referenced to the dissipation $\Gamma$. Colored areas are obtained by solving Equation~\ref{eq: 1}. Black curves are Boltzmann distributions for the potential $V_{\mathrm{app}}$. Inset: Phase diagram of the Kerr-nonlinear cavity, with colored dots indicating the values of $A/\sqrt{\Gamma}$ and $\Delta/\Gamma$ considered in the main panel. (b) Deviation $\epsilon$ from Boltzmann statistics as defined by Eq.~\ref{eq: 8}, for the same values of $A/\sqrt{\Gamma}$ and $\Delta/\Gamma$ considered in (a). $D/\sqrt{\Gamma}=1/2$ in the main panel of (a) and in (b). (c) Ratio of off-diagonal to diagonal parts of the first matrix in the right hand side of Eq.~\ref{eq: 2}. This ratio quantifies the mutual coupling between the field components $\alpha_R$ and $\alpha_I$, and also the relative strength of non-conservative and conservative forces. The two gray curves correspond to the two states in the bistability. The black curve is the mean ratio. The orange cross marks the boundary between weak and strong coupling between $\alpha_R$ and $\alpha_I$. (d) Same as in (b) but for two additional standard deviations of the noise $D$. }
\end{figure}

In this section we quantitatively compare predictions based on $V_{\mathrm{app}}$ to numerical simulations of the full system. One one hand, we calculate distributions of field amplitudes $\sqrt{N}$ for the full system along the path $\ell$. We call those distributions $P_{\mathrm{full}}(\sqrt{N})$. We obtained $P_{\mathrm{full}}(\sqrt{N})$ by numerically solving Eq.~\ref{eq: 1} with $U/\Gamma=0.01$ and $D/\sqrt{\Gamma} =1/2$. We evolved the system for a time $\Gamma t=10^6$ and ran simulations for 16 different realizations of the noise $\zeta(t)$. The resultant $P_{\mathrm{full}}(\sqrt{N})$ are shown in Fig.~\ref{fig:5} as areas of different color for different driving conditions ($A/\sqrt{\Gamma}, \Delta/\Gamma $). On the other hand, we calculate Boltzmann distributions by inserting the approximate potential $V_{\mathrm{app}}$ and the effective temperature $T=D^2/2k_B$ in Equation~\ref{eq: 5}. The Boltzmann distributions, which we call $P_{\mathrm{Bol}}(\sqrt{N})$, are shown as black curves on top of $P_{\mathrm{full}}(\sqrt{N})$ in Fig.~\ref{fig:5}.

For small $A/\sqrt{\Gamma}$ and $\Delta/\Gamma$, meaning small $\Omega$, $P_{\mathrm{full}}(\sqrt{N})$ is in very good agreement with $P_{\mathrm{Bol}}(\sqrt{N})$. This is expected based on the results in Fig.~\ref{fig:2}(b), where $\Omega=0$. However, as $\Omega$ increases, $P_{\mathrm{full}}(\sqrt{N})$ increasingly deviates from $P_{\mathrm{Bol}}(\sqrt{N})$. We quantify the difference between the two distributions via the overlap integral
\begin{align}
    \epsilon = \frac{1}{2}\int|P_{\mathrm{full}}-P_{\mathrm{Bol}}|d\sqrt{N}.
    \label{eq: 8}
\end{align}
\noindent $\epsilon$ is 0 when $P_{\mathrm{full}}(\sqrt{N})$ exactly matches $P_{\mathrm{Bol}}(\sqrt{N})$, and it is $1$ when there is zero overlap between the two distributions.

Figure~\ref{fig:5}(b) shows that $\epsilon$ is a non-monotonic function of the distance to the critical point, controlled via the value of $\Delta/\Gamma$. A large $\epsilon$ is presumably the result of non-conservative forces, which are absent in the Boltzmann distribution taking into account $V_{\mathrm{app}}$ only. We tested this hypothesis by calculating the ratio of $|\Omega|^2$ to $\Gamma^2/4$. As Eq.~\ref{eq: 2} evidences, this ratio is proportional to the ratio of non-gradient (i.e., non-conservative) to gradient forces. The calculation was done for driving conditions along the dashed line in Fig.~\ref{fig:3}(b), crossing the critical point. The result is shown in Fig.~\ref{fig:5}(c). We focus on the `mean' ratio [black curve in Fig.~\ref{fig:5}(c)] because there are two distinct values of $|\Omega|^2$ [dashed gray curves in Fig.~\ref{fig:5}(c)] in the bistable regime, and the distributions $P_{\mathrm{full}}(\sqrt{N})$ reflect both values. Remarkably, the mean ratio $4|\Omega|^2 / \Gamma^2$ displays a very similar dependence on $\Delta/\Gamma$ as $\epsilon$. This indicates that deviations from Boltzmann statistics are indeed associated with the non-gradient force, which is proportional to the mutual coupling $\Omega$ between the field components $\alpha_R$ and $\alpha_I$.

Next we assess whether the noise strength affects the overlap between $P_{\mathrm{full}}(\sqrt{N})$ and $P_{\mathrm{Bol}}(\sqrt{N})$. In Fig.~\ref{fig:5}(d) we compare $\epsilon$ as a function of $\Delta/\Gamma$ for three different values of $D/\sqrt{\Gamma}$. For small $\Delta/\Gamma$, $\epsilon$ is roughly independent of $D/\sqrt{\Gamma}$. In that regime, the non-gradient force is relatively weak, so the insensitivity of $\epsilon$ to $D$ is not so surprising. Interestingly, for $\Delta/\Gamma \gtrsim 0.9$ we observe very significant differences in $\epsilon$ for the three $D/\sqrt{\Gamma}$. For example, for $\Delta/\Gamma =0.97$, $\epsilon$ is $\sim5$ times larger for $D/\sqrt{\Gamma}=1/2$ than for $D/\sqrt{\Gamma}=1$. For a larger $\Delta/\Gamma$, $\epsilon$ is further reduced as $D/\sqrt{\Gamma}$ increases. This suggests that the stochastic force effectively suppresses the effects of the non-gradient force. Thus, the distribution of light in the bistable cavity resembles more the Boltzmann distribution of an equilibrium system in a scalar double well potential for strong noise. Typically, the correspondence between non-equilibrium and equilibrium systems hinges on the similarity (or equivalence) of the deterministic equations of motion governing the behavior of the two systems. Here, in contrast, we have found that the accuracy of the correspondence also depends on the noise strength.

While $P_{\mathrm{full}}(\sqrt{N})$ strongly deviates from $P_{\mathrm{Bol}}(\sqrt{N})$ deep in the bistability regime, the deviation is quite small close to the critical point. Recall that the critical point is at $\Delta/\Gamma = \sqrt{3}/2 \approx 0.87$, and notice in Fig.~\ref{fig:5}(d) that $\epsilon \approx 0.05$ around $\Delta/\Gamma = 0.9$. The small ($\sim 5\%$) deviation from equilibrium behavior justifies our claim that $V_{\mathrm{app}}$ approximately captures the dynamics of the full system. This is an important and convenient result because most of the interesting physics occurs near the critical point. Our results therefore indicate that stochastic light in a bistable optical cavity close to criticality can be considered approximately equivalent to a Brownian particle in a double well potential.

\section{Phase Transition for the Fluctuations} \label{section:fluctuations}
We have previously shown that the transition from Boltzmann to non-Boltzmann statistics of light in the nonlinear cavity is associated with an increased coupling between the field components $\alpha_{R, I}$. Here we show that $\alpha_{R, I}$ actually transition from weak to strong coupling concomitantly with a phase transition for the fluctuations. This phase transition conveys qualitative changes to the phase portrait of the system, enabling us to understand the approximate validity of the potential $V_{\mathrm{app}}$ within a restricted parameter regime.

Consider the effect of adding a small fluctuation $\delta\alpha = \delta\alpha_R + i\delta\alpha_I$ to the light field, i.e., let $\alpha\rightarrow \alpha+\delta\alpha$ in Eq.~\ref{eq: 1}. By only retaining terms that are linear in the fluctuations, we get the following matrix equation of motion for the fluctuations:

\begin{align}
\begin{pmatrix}
 \delta\Dot{\alpha}_R \\ \delta\Dot{\alpha}_I
\end{pmatrix}=
   \begin{pmatrix}
     -\frac{\Gamma}{2}+2U\alpha_R\alpha_I&    U(\alpha_R^2+3\alpha_I^2) -\Delta   \\
   \Delta -U(3\alpha_R^2  + \alpha_I^2)      & -\frac{\Gamma}{2}-2U\alpha_R\alpha_I  \\
\end{pmatrix}
\begin{pmatrix}
 \delta\alpha_R \\ \delta\alpha_I
\end{pmatrix}.
\label{eq: 9}
\end{align}

\noindent Equation~\ref{eq: 9} has solutions of the form

\begin{equation}
      \vec{ \delta\alpha} =\vec{\eta}e^{\lambda t}
    \label{eq: 10}
\end{equation}

\noindent where $\vec{\eta}$ are the eigenvectors and $\lambda$ the eigenvalues of the $2\times2$ matrix in Eq.~\ref{eq: 9}.
The eigenvalues

\begin{equation}
      \lambda_{\pm} = -\Gamma/2 \pm \sqrt{G(U,\Delta,N)}
    \label{eq: 11}
\end{equation}

\noindent comprise the spectrum of the fluctuations. The function $G = -(\Delta-UN)(\Delta-3UN)$ determines the stability of the fixed points, and the validity range of the potential $V_{\mathrm{app}}$ as explained next.

\begin{figure}[t]
	\includegraphics[width=\columnwidth]{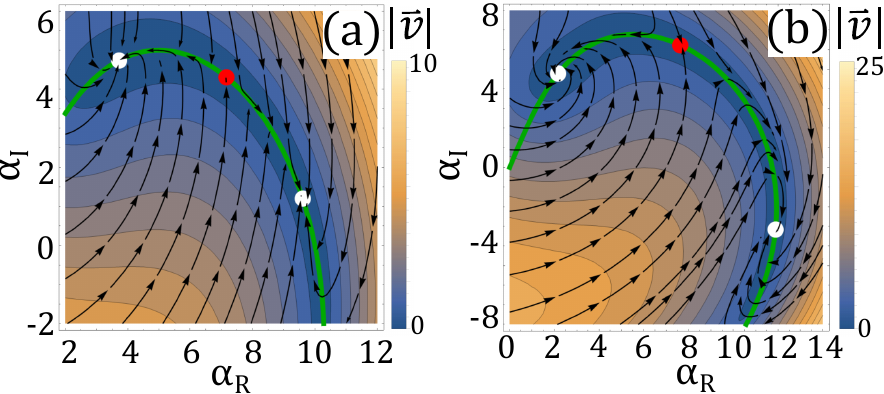}
 	\caption{\label{fig: 6}(a,b) Phase portrait of the Kerr-nonlinear cavity evaluated at the driving amplitudes and detunings indicated in Fig.~\ref{fig:7}(a). Arrows and color have the same meaning as in Fig.~\ref{fig:2}(a). White dots are stable fixed points, red dots are unstable fixed points. Green curves indicate the path $\ell$ where $\Dot{\alpha}_I=0$ and the potential $V_{\mathrm{app}}$ is defined. $V_{\mathrm{app}}$ works well in (a) where there is approximate gradient flow behavior, but not in (b) where non-conservative forces dominate and lead to spiraling orbits around the stable fixed points.}
\end{figure}

\begin{figure}
    \centering
    \includegraphics[width=\columnwidth]{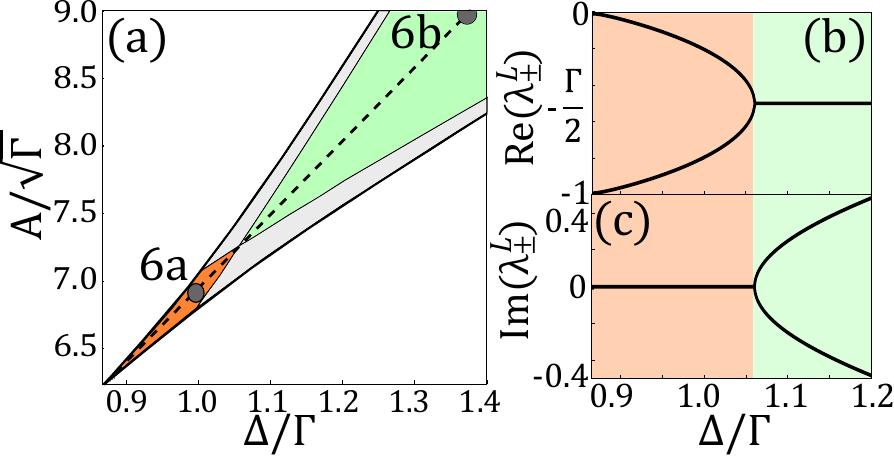}
    	\caption{\label{fig:7} (a) Phase diagram of a Kerr resonator, as in Fig.~\ref{fig:3}(b) but for a reduced range of driving amplitude $A$ and detuning $\Delta$ both referenced to the dissipation $\Gamma$. The orange region close to the critical point is where the eigenvalues $\lambda_{\pm}$ in Eq.~\ref{eq: 11} are purely real, the field components $\alpha_R$ and $\alpha_I$ are weakly coupled, and there is a saddle-sink connection between the fixed points. This is the region where $V_{\mathrm{app}}$ works well. The green region is where both eigenvalues of both states have a non-zero imaginary part. In the gray region, one of the two states has eigenvalues with non-zero imaginary part. $V_{\mathrm{app}}$ fails to properly capture system's dynamics in the gray and green regions. (b) and (c) are real and imaginary parts of the eigenvalues of the low-density state along the dashed line in (a). Orange and green regions have the same meaning as in (a). Real and imaginary parts of the eigenvalues coalesce at $\Delta/\Gamma=1.06$, which resembles an exceptional point.}
\end{figure}

Figure~\ref{fig: 6} shows how the force field $\vec{v}= (\Dot{\alpha}_R,\Dot{\alpha}_I)$ experienced by light in the nonlinear cavity is shaped by the function $G$. Figures~\ref{fig: 6}(a) and ~\ref{fig: 6}(b) correspond to positive and negative $G$, respectively. Figure~\ref{fig: 6}(a) was obtained for $A/\sqrt{\Gamma}=6.95$ and $\Delta/\Gamma=1$, which places the cavity within the bistability regime and close to the critical point [see Fig.~\ref{fig:3}(b) or Fig.~\ref{fig:7}(a)]. Stable and unstable fixed points are represented by white and red dots, respectively. Note that the unstable fixed point has purely real eigenvalues with opposite sign, and it is therefore always a saddle point. The green curve is the path $\ell$ (where $\Dot{\alpha}_I=0$) along which we evaluated $V_{\mathrm{app}}$. Notice how the vectors $\vec{v}$ all point to the path $\ell$, which connects the unstable (saddle) point and stable fixed points. If the system is taken away from $\ell$ by the stochastic force, the deterministic force will ensure that it returns to $\ell$. Therefore, the system almost behaves as a gradient flow system in one dimension. This elucidates why the one-dimensional potential $V_{\mathrm{app}}$ approximately captures the full system's dynamics close to the critical point. We can also understand this behavior based on the spectrum of fluctuations. For driving conditions giving $G>0$, the eigenvalues $\lambda_{\pm}$ turn out to be real negative numbers. This makes the fluctuations overdamped and the stable fixed points are sinks. Consequently, we observe gradient-flow-like behavior and the bistable cavity approximately behaves like an overdamped oscillator in a one-dimensional scalar potential.

The physics is different for $G<0$. For example, Fig.~\ref{fig: 6}(b) shows the phase portrait for $A/\sqrt{\Gamma}=9$ and $\Delta/\Gamma=1.38$, which places the cavity within the bistability regime but further away from the critical point than in Fig.~\ref{fig: 6}(a). The change of parameters has transformed the stable sinks into stable foci. Each stable focus is evidenced by a spiraling force field around a white dot in Fig.~\ref{fig: 6}(b). The spiraling force field implies that a fluctuation in one field component ($\alpha_R$ or $\alpha_I$) couples to the other component. When this coupling is strong, the non-gradient force dominates, the fluctuations are no longer overdamped, and any perturbation causes the field to stabilize at a new orbit in the two-dimensional force field. Clearly, a one-dimensional potential cannot faithfully capture the full system's dynamics in this regime.

The behavior deduced from the phase portraits in Figs.~\ref{fig: 6}(a,b) can be further elucidated by considering the spectrum of fluctuations. For this purpose, we calculated the eigenvalues $\lambda_{\pm}$ in Equation~\ref{eq: 11}, focusing on the parameter range resulting in bistability. Based on these results, we distinguish three regimes depending on the driving conditions. These regimes are illustrated as areas of different colors in Fig.~\ref{fig:7}(a). In the orange region, all eigenvalues $\lambda_{\pm}$ are purely real and negative for both bistable states. Consequently, the fluctuations are overdamped and there is a saddle-sink connection between the fixed points. In that regime, $\ell$ (defined by $\Dot{\alpha}_I=0$) closely follows the most probable path between the stable fixed points, and the dynamics is approximately captured by the one-dimensional potential $V_{\mathrm{app}}$ defined along $\ell$. In contrast, the dynamics in the gray and the green regions cannot be described in terms of $V_{\mathrm{app}}$ only. This is because the eigenvalues are imaginary (oscillating fluctuations) for one of the steady states in the gray region, and for both states in the green region.

In Fig.~\ref{fig:7}(b) and ~\ref{fig:7}(c) we plot the real and imaginary parts of $\lambda_{\pm}$, respectively, for the low photon density bistable state. We plot the eigenvalues as a function of $\Delta/\Gamma$ while also varying $A/\sqrt{\Gamma}$, thereby keeping the system along the dashed line in Fig.~\ref{fig:7}(a). The orange region indicates the parameter range for which the fluctuations are overdamped and effectively decoupled. The green region indicates the parameter range for which the fluctuations are oscillatory and strongly coupled. Interestingly, real and imaginary parts of both eigenvalues coalesce at the boundary between these two regions. To the left (resp. right) side of this coalescence point, the real (resp. imaginary) parts of $\lambda$ split while the imaginary  (resp. real) parts cross. This is the typical behavior of the eigenvalues of a non-Hermitian Hamiltonian describing two linearly-coupled linear modes~\cite{Moiseyev11, Heiss12, Rodriguez16EJP, Miri19, Soriente21}. Thus, the degeneracy point at $\Delta/\Gamma = 1.06$ acts as an exceptional point for the fluctuations.

The exceptional point for the fluctuations separates the regimes in which a bistable cavity can or cannot be approximately described by the potential $V_{\mathrm{app}}$. This is an interesting analogy to the non-Hermitian physics of coupled modes. There, the exceptional point defines the boundary between weak and strong coupling between the two modes~\cite{Rodriguez16EJP}. The similarity is even more striking when we consider that $\Delta/\Gamma = 1.06$ also corresponds to point at which the mutual coupling between field components $\alpha_{R,I}$ transitions from weak to strong. Indeed, for $\Delta/\Gamma = 1.06$ we have $4|\Omega|^2 \approx \Gamma^2$, which corresponds to the boundary between weak and strong coupling between the field components $\alpha_{R,I}$. The point $4|\Omega|^2 = \Gamma^2$ is indicated by an orange cross in Fig.~\ref{fig:5}(c). The large values of $\epsilon$ above this point in Fig.~\ref{fig:5}(c) indicate that the large deviation from Boltzmann statistics is indeed related to a phase transition for the fluctuations occurring at the exceptional point.

\section{Conclusions and Perspectives} \label{section:conclusion}
To summarize, we have shown that stochastic light in a coherently-driven nonlinear optical cavity is mathematically equivalent to two coupled overdamped Langevin oscillators. Whether a scalar potential can fully capture the system's dynamics or not depends on the driving amplitude and frequency. These parameters determine the mutual coupling between the Langevin oscillators comprising the complex light field. For weak on-resonance driving, there is an exact correspondence: The dissipation and the driving amplitude define a scalar potential, the noise variance defines an effective temperature, and the distribution of light in the cavity satisfies Boltzmann statistics. This effective equilibrium behavior is approximately valid for moderately strong non-resonant driving in the bistable regime, but breaks down deep in the bistability regime. The relevance of these results stems from the fact that the overdamped Langevin oscillator is a cornerstone of statistical physics and stochastic thermodynamics~\cite{Sekimoto98, Jarzynski11, Seifert12, Ciliberto17}. Numerous important results about fluctuations of thermodynamic quantities, the efficiency of stochastic engines, and the precision of information-processing systems, have emerged from understanding Langevin dynamics in the overdamped limit~\cite{Sekimoto98, Seifert12, Ciliberto17, Chernyak06,  Pan18}. By defining an effective temperature and a potential for light in an optical cavity, our work provides a first step towards understanding resonant optical systems within the framework of stochastic thermodynamics. Thermodynamic quantities like heat and work still need to be defined, and we view that as an important future research direction. We foresee exciting discoveries in that direction, enabled by the ability to exactly or approximately describe resonant optical systems as Brownian particles in scalar potentials.

\vspace{5mm}

\section*{Acknowledgments}
\noindent This work is part of the research programme of the Netherlands Organisation for Scientific Research (NWO). We thank Nicola Carlon Zambon, Kevin Peters, Pieter Rein ten Wolde, and Martin van Hecke for stimulating discussions. S.R.K.R. acknowledges an ERC Starting Grant with project number 85269.

%

\end{document}